# МОДУЛЬНАЯ ТЕХНОЛОГИЯ РАЗРАБОТКИ РАСШИРЕНИЙ САПР. АКСОНОМЕТРИЧЕСКИЕ СХЕМЫ ТРУБОПРОВОДОВ. ПАРАМЕТРИЧЕСКОЕ ПРЕДСТАВЛЕНИЕ

Мигунов В.В., Кафиятуллов Р.Р., Сафин И.Т.
ЦЭСИ РТ, Казань
vmigunov@csp.kazan.ru

Рассматривается применение модульной технологии разработки проблемно-ориентированных расширений САПР к задаче автоматизации подготовки аксонометрических схем трубопроводных систем на примере программной системы TechnoCAD GlassX. Выявлено единство состава схем для специальных технологических трубопроводов, систем водопровода и канализации, отопления, теплоснабжения, вентиляции, кондиционирования воздуха. Приводится структурированное параметрическое представление схем, включая свойства объектов и их связи, общие установки и установки по умолчанию. Рассмотрены специальные связи совместимости.

Настоящая работа посвящена применению модульной технологии разработки проблемно-ориентированных расширений САПР реконструкции предприятия, общие положения которой изложены в [1]. Объект приложения технологии - автоматизация подготовки аксонометрических схем трубопроводных систем, выполненная в САПР TechnoCAD GlassX.

Анализ системы проектной документации для строительства показывает, что аксонометрические схемы трубопроводных систем (АСТС) включаются в состав следующих рабочих чертежей: чертежи специальных технологических трубопроводов [2], схемы систем водопровода и канализации [3], схемы систем отопления, теплоснабжения, вентиляции, кондиционирования воздуха [4]. Схемы в этом списке выполняются в различных проекциях, но все они требуются для реконструкции предприятия и допускают единое представление в чертеже в виде совокупности следующих элементов (не обязательно всех одновременно):

- линии, изображающие трубопровод, включая разрывы в виде серии точек или двух волн обрыва;
- условные графические обозначения опор, элементов трубопроводов и трубопроводной арматуры;
- буквенные обозначения разрывов труб;
- текстовые сведения о маркировке и диаметрах труб, о типе и размерах опор, об обозначениях элементов трубопроводов и арматуры;
- текстовые сведения об уклонах труб;
- другая текстовая информация, более общего характера, относящаяся к трубам, опорам, элементам трубопроводов и трубопроводной арматуре;
- размеры, характеризующие длины труб, положение поворотов и концов труб, положение опор и арматуры на трубах;
- отметки высоты поворотов и концов труб, высоты расположения опор и арматуры на трубах;
- оси плана строительной подосновы;
- вспомогательное изображение аксонометрических осей координат (картинка с осями);
- изображения аппаратов, в которые входят трубы.

Поддержка различных проекций обеспечивается установкой текущей проекции, которая удобна и для рабочих просмотров АСТС с разных позиций. Изображения аппаратов вычерчиваются относительно самостоятельно и слабо связаны с остальной частью схемы, в параметрическое представление АСТС они не включаются. Параметрическое представление объектов и их связей в АСТС содержит следующие списки, в описании которых отражены не только связи принадлежности, но и более сложные ограничивающие связи (условия совместимости). Кроме того, учтена необходимость специфицирования АСТС.

*Точки*

Список пространственных точек, являющихся концами осей прямых труб (истинными, без учета разрывов на чертеже). Совпадающие точки объединяются в одну. Точки – единственные объекты схемы, задающие ее пространственную конфигурацию.

*Трубы*

В списке труб каждая труба задается ссылкой на точку начала и точку конца ее оси из списка точек, цветом и типом линии. Запрещаются трубы нулевой длины и наложения труб.

*Соединения труб*

Список парных соединений труб между собой содержит для каждого соединения ссылки на две трубы и вид соединения – встык или радиус сопряжения. На одну пару труб может быть не более одного соединения. Соединения не применяются при автоматическом определении связной линии трубопровода, для связности достаточно, чтобы трубы шли из одной точки. Цвет и тип линии сопряжения труб задается цветом и типом линии самих труб (любой из них).

*Смещения*

В АСТС поддерживается список смещений, возникающих из-за линий разрыва. В месте разрывов трубопровод изображается точками (при растяжении) или двумя линиями обрыва (при сжатии) и не в масштабе, а переносится вдоль оси трубы для исключения затенения или для сокращения видимых размеров схемы, как это показано в ГОСТ 21.601-79 (Черт.8). Это означает, что с одной стороны от заданной плоскости все точки и связанные с ними концы труб смещаются по нормали к этой плоскости на заданную величину. Каждому смещению соответствует идентифицирующая его буква, орт направления из несмещаемой части схемы в смещаемую и положительная (для растяжения) либо отрицательная (для сжатия) величина смещения.

Различаются общие и местные смещения. Общее определяется координатой плоскости, разделяющей пространство на две части, и автоматически порождает линии разрыва на трубах, пересекающих ее. Все трубы и размерные линии, пересекаемые плоскостью общего смещения, могут идти только по нормали к ней. Иначе возникают недопустимые разрывы не вдоль их осей. Местное смещение определяется совокупностью установленных разрывов труб, каждый из которых смещает свою ветвь трубопровода. Совокупность установленных на трубы разрывов для одного местного смещения должна позволять пересечь все эти разрывы одной поверхностью без самопересечений, которая разделит все пространство на две части таким образом, что с одной стороны от поверхности на все точки схемы смещение действует, а с другой стороны – не действует.

Буквенные обозначения различных смещений должны быть различными. При добавлении нового смещения его буквенное обозначение генерируется по возрастанию в последовательности а, б, в, ...

*Линии разрыва*

В списке линий разрыва для каждой из них хранятся ссылки на трубу из списка труб и на смещение из списка смещений; длина на чертеже, изображаемая точками;

смещение середины разрыва от плоскости разрыва вдоль вектора смещения для общих смещений либо положение на трубе разрыва для местных смещений; смещение буквенного обозначения разрыва в чертеже от концов разрыва вдоль оси трубы и по нормали к ней. Буквенные обозначения линий разрыва берутся из их смещений и располагаются симметрично относительно середины линии разрыва. Для разрывов, изображающих сжатие, допускается вариант изображения их двумя волнами обрыва, диаметр которых задается общим для всей схемы. Линии разрыва при растяжении схемы всегда изображаются точками и располагаются вдоль всей раздвинутой части труб, для них свойства длины и смещения середины от плоскости разрыва не играют роли.

*Внутренняя библиотека обозначений*

Собственная внутренняя библиотека условных графических обозначений арматуры, опор и элементов трубопроводов, используемых в данном модуле АСТС (блоков), содержит графические обозначения в их исходном двумерном состоянии, то есть двумерную графическую часть, сведения о возможной привязке (осевой, угловой или тройниковой), сведения о вырезаемых длинах труб при установке обозначения на трубы, признаки симметрии геометрии и привязок относительно оси привязки и нормали к ней, коэффициент растяжения. Привязки и вырезаемые длины используются для автоматического расположения блоков на трубах с удалением из труб закрываемых блоком частей. Признаки симметрии применяются для сокращения числа перебираемых проектировщиком вариантов пространственной ориентации блока при установке его на трубу. Коэффициент растяжения относительно подключенной графической библиотеки используется для коррекции размеров обозначений в схеме.

*Блоки*

Список установленных на трубы условных графических обозначений арматуры, опор и элементов трубопроводов для каждого из них содержит ссылки в собственную библиотеку и на трубу из списка труб, две необязательных ссылки на вторую и третью привязанные трубы (для угловой и тройниковой привязок), цвет и тип линии, расстояние от точки привязки блока до начала трубы, два признака ориентации блока в пространстве. Вариантов этой ориентации не более 16, поскольку ГОСТ 2.317-69 требует располагать выносные линии размеров вдоль аксонометрических осей координат, а отходящие от условных графических обозначений выносные линии должны находиться в плоскости обозначения. Первый признак задает, идет ли ось X+ блока от начала трубы к концу или обратно. Второй – задает ось в пространстве, с которой образует острый угол ось Y+ блока (нормаль к его оси). Этой осью могут быть X+, X–, Y+, Y–, Z+, Z–, а для угловых и тройниковых блоков – также и одна из привязанных труб, не идущих по оси блока.

*Тексты*

Список текстов для каждого из них содержит собственно многострочный текст, ссылку на главную сноску в одном из списков сносок, установку шрифта, шаг строк, цвет, двумерный вектор смещения в чертеже (на Натуре) от точки указания главной сноски до начала текста. От одного текста идут одна или более сносок к трубам и блокам, одна из них считается главной. Соответствующая труба или блок определяют положение текста на схеме. Остальные сноски считаются неглавными. Для текстов, обозначающих уклоны труб, используются специальные символы "Уклон влево" и "Уклон вправо", "Градус". Для указания диаметров труб используется специальный символ "Диаметр". Если главная сноска текста указывает на трубу и текст содержит символ уклона, то сохраняется также и формат значения уклона (угол, отношение, процент). В этом случае при всяких изменениях реального уклона трубы новое значение уклона проставляется в текст.

*Сноски от текстов к трубам*

Список сносок от текстов к трубам для каждой сноски содержит ссылки на текст из списка текстов и на трубу из списка труб, координату на трубе от ее начала к концу в мм Натуры. Цвет сноски берется из цвета ее текста;

*Сноски от текстов к блокам*

Список сносок от текстов к блокам для каждой сноски содержит ссылки на текст из списка текстов и на блок из списка блоков, точку положения конца сноски на библиотечном двумерном изображении обозначения (на Бумаге). Цвет сноски берется из цвета ее текста.

*Позиционные обозначения*

Список позиционных обозначений для каждого из них содержит ссылку на трубу или блок, точку положения на нем конца сноски, от 1 до 8 (в соответствии с количеством позиций) ссылок на специфицирующие свойства, установку шрифта, шаг строк, цвет, двумерный вектор смещения в чертеже (на Натуре) от точки указания сноски до начала текста обозначения, вариант проведения сноски от начала или от конца полки, а также признак видимости позиционного обозначения (невидимое позиционное обозначение не будет вычерчиваться только в том случае, если не установлена видимость скрытых позиционных обозначений). В зависимости от установки автонумерации позиции задаются вручную или автоматически перенумеровываются, плотно занимая диапазон от 1 и далее, в том числе при удалении позиционных обозначений. Часть позиций соответствует невидимым на схеме, но представленным в спецификации элементам – крепежным деталям и, возможно, прокладкам.

*Специфицирующие свойства*

Список специфицирующие свойства элементов трубопроводов содержит для каждой позиции комплект свойств, зависящий от специфицируемого объекта (для блоков, в отличие от труб, имеется дополнительное свойство "Кол." – количество) и установки, определяющей, какой набор специфицирующих свойств хранится и задается:

только для 6-тиграфной спецификации – комплект включает свойства "Поз." (позиция, только если позиции задаются вручную), "Обозначение", "Наименование", "Масса ед." (масса единицы, кг), "Примеч." (примечание);

расширенный (и для 6-ти графной и для заказной спецификаций) – комплект дополнительно включает свойства "Тип, марка" (тип, марка, обозначение документа, опросного листа), "Наименование и техническая характеристика", "ЕдИзм" (наименование единиц измерения), "Завод-изготовитель" (наименование завода-изготовителя), "Код оборудов." (код оборудования, изделия, материала).

Эти свойства на схеме невидимы и просматриваются только при их правке либо при генерации спецификации. Источником свойств является выбор в электронных каталогах или прямой ввод с клавиатуры. При автоматическом специфицировании суммарная длина труб или суммарное количество блоков определяется автоматически.

*Размеры*

Каждый размер в их списке считается цепным, обычный линейный является его частным случаем. Размер задается списком точек, из которых идут выносные линии (это могут быть точки из списка пространственных точек либо точки привязки установленных блоков), ориентацией выносных линий (оси X, Y или Z), ориентацией размерной линии (оси X, Y, Z либо ссылка на трубу, если размерная параллельна ее оси), смещением размерной от первой точки и смещением текстов размера от размерной линии (оба в мм Бумаги). Индивидуальных установок шрифта и стрелок, выхода выносных линий за размерную, цвета и точности подсчета размеры не имеют, они подчиняются общим установкам для размеров схемы. В тексты размеров автоматически проставляются имеющиеся реальные значения этих размеров. Также автоматически стрелки заменяются засечками и стрелки внутрь меняются на стрелки наружу, если не хватает места между выносными для помещения двух стрелок внутри.

Набор точек начала выносных линий определяет допустимые варианты ориентации выносных и размерной. Точки должны лежать в одной плоскости, среди них не должно быть совпадающих. Во всех случаях, если часть точек подвергается смещению при разрыве, а часть – нет, смещение может проходить только в плоскости точек. Также всегда направление выносных линий не должно совпадать с направлением размерной.

Если точки не лежат на одной пространственной оси, то их плоскость должна быть параллельна одной из координатных плоскостей. Не должно быть смещений, которые сдвигают часть точек относительно другой их части и выводят часть точек из этой плоскости. Ни размерная, ни выносные не могут идти перпендикулярно плоскости точек, а в плоскости точек остаются два варианта чередования выносных и размерной по двум осям.

Если же точки лежат на одной оси в пространстве, то либо это должна быть ось координат, либо ось имеющейся трубы. Не должно быть смещений, которые сдвигают часть точек относительно другой их части и выводят часть точек с этой оси. Если это ось координат, то размерная может идти только вдоль нее, а выносные по двум оставшимся осям координат. Если это ось трубы, не параллельная ни одной из координатных плоскостей, то размерная может идти только по ней, а выносные – по всем трем осям. Если это ось трубы, параллельная одной из координатных плоскостей, размерные и выносные не могут идти по перпендикулярной к ней оси координат. Допускается три варианта ориентации размерной: по трубе и по двум осям координат в этой плоскости. Если размерная идет по оси трубы, выносная может идти по любой из двух осей координат в этой плоскости. Если размерная идет по одной из этих осей координат – выносная идет по другой.

*Отметки высоты*

Список отметок высоты для каждой отметки содержит сведения о точке, из которой идет выносная линия, аксонометрические оси направления выносных линий (X или Y) и полки (X+, X–, Y+, Y–); смещение точки указания стрелкой вдоль выносной линии, смещение полки вверх от выносной (положительное или отрицательное); тип линии стрелки отметки высоты. Своего цвета отметки не имеют, применяется установка их общего в схеме цвета. В качестве точек начала выносных выступают либо точки на трубах (включая их концы), либо точки привязки условных графических обозначений к трубам.

*Обозначения уклонов*

Список обозначений уклонов для каждого обозначения содержит сведения о точке на трубе и расстояние (положительное или отрицательное) от нее до середины обозначения, формат и точность представления уклонов. Все обозначения уклонов отображаются одним цветом, также они имеют общую установку шрифта и общий размер стрелки.

*Строительные оси*

Два списка групп строительных осей (соответственно осей X и осей Y) содержат для каждой из групп осей их число в группе и шаг. Информация об осях берется из строительной подосновы путем выбора в чертеже или чтения из файла параметров и всегда содержит полный комплект сведений об осях подосновы. Видимы на схеме лишь оси, указываемые пользователем. Своего цвета группы строительных осей не имеют, он берется из установок схемы. Совокупность всех строительных осей (и по X, и по Y) является одним объектом схемы и может удаляться только целиком. Имеющиеся в схеме группы осей можно исправлять, удалять и добавлять, но заново ввести их можно только путем импорта.

*Установки схемы*

Установки АСТС делятся на установки объектов и установки режима работы. Установки объектов могут действовать по умолчанию либо на всю схему. Действие по умолчанию означает, что установки применяются при добавлении объекта в АСТС.

*Установки объектов*
- цвет и тип линии труб по умолчанию;
- вид соединения (встык или радиус сопряжения в мм Натуры) по умолчанию;
- для линий разрыва: длина и вектор смещения буквенных обозначений в мм Бумаги по умолчанию; общие для схемы шаг точек в мм Бумаги и установка шрифта буквенных обозначений;
- коэффициент растяжения условных обозначений по умолчанию;
- цвет и тип линии блоков по умолчанию;
- установка шрифта, цвет, сдвиг строк по умолчанию, вариант проведения сноски от начала или от конца полки по умолчанию в текстах, наличие второй полки при двух строках;
- установка шрифта, цвет, сдвиг строк, вариант проведения сноски от начала или от конца полки по умолчанию в позиционных обозначениях;
- общие для схемы: шрифт, установка стрелок и точности подсчета размеров, цвет размеров; смещение текстов размера от размерной линии в мм Бумаги по умолчанию;
- тип линии, направление выносной и полки, смещение точки указания стрелкой вдоль выносной линии в мм Бумаги, смещение полки от выносной в мм Бумаги в отметках высоты по умолчанию; общие для схемы шрифт текстов, длина стрелки в мм Бумаги и цвет отметок высоты;
- смещение обозначений уклонов от трубы в мм Бумаги, формат и точность представления уклонов по умолчанию; общие для схемы шрифт текстов, размер стрелки (длина и размах крыльев в мм Бумаги) и цвет обозначений уклонов;
- большое количество общих для схемы установок построения сетки строительных осей. Признак обозначения цифрами X осей, координата Z плоскости сетки, смещение по Z в мм Бумаги продолжений осей, загибающихся вверх или вниз, список видимых осей, сдвиги размерной линии от сетки наружу в мм Бумаги вдоль осей X и вдоль осей Y, длины продолжений за сетку в мм Бумаги первых осей X и Y, первый номер и первая буква в обозначениях осей, признаки наличия размеров между крайними осями для осей X и Y, ориентация сетки относительно АСТС (ось X идет в положительном направлении или нет, то же для оси Y), расположение номерных и буквенных обозначений (у первой или у последней из видимых осей другого направления), цвет;
- количество позиций по умолчанию в позиционном обозначении фланцевого соединения.

*Установки режима работы*
- длина разрывов труб АСТС, показывающих затенение их друг другом;
- текущее имя файла с комплектом параметров АСТС (последнее выбранное);

- вариант текущей аксонометрической или обычной проекции схемы;
- выборка объектов из схемы, то есть слой схемы по высоте;
- признаки видимости различных объектов АСТС. Все сноски видны или не видны вместе с текстами. Кроме самих объектов, здесь устанавливается также видимость затенений труб друг другом, видимость буквенных обозначений разрывов труб, видимость полностью закрытых блоками и сопряжениями труб, а также видимость скрытых позиционных обозначений. Управление видимостью объектов обеспечивает наглядность и быстродействие при подготовке схем;
- установки фильтрации данных в электронных каталогах для специфицирования схем: рабочие температура и давление среды;
- признак автонумерации позиций позиционных обозначений. Если установлен, то позиции автоматически перенумеровываются, плотно занимая диапазон от 1 и далее, в том числе при удалении позиционных обозначений, иначе позиции задаются вручную;
- признак, определяющий, какой набор специфицирующих свойств хранится и задается. Если установлен, то только для 6-тиграфной спецификации, не установлен – расширенный (и для 6-ти графной и для заказной спецификаций).

Варианты используемых проекций не ограничиваются задаваемыми в ГОСТ 21.401-88, ГОСТ 21.601-79, 21.602-79 фронтальной диметрической и изометрической проекциями. Обеспечиваются все допустимые по ГОСТ 2.317-69 13 аксонометрических проекций, а также более простые виды по ГОСТ 2.305-68, дополнительные 6 косоугольных фронтальных проекций (3 изометрические и 3 диметрические) с направлением оси X горизонтально вправо и направлением оси Y в первом квадранте плюс абстрактное проецирование вдоль пространственных осей, задаваемых пользователем при просмотре АСТС в режиме облета.

Кроме проекции как таковой, при создании изображения может задаваться и пространственный слой. Например, часть схемы, попавшая в один этаж, то есть между двумя отметками высоты. В этом случае из схемы отбирается часть объектов по следующим правилам: трубы – если задевают слой, соединения – если соединяют отобранные трубы, линии разрыва – вместе с трубами, блоки – если точки привязки находятся внутри слоя, тексты и сноски – если указывают на выбранные трубы и блоки, размеры – если хоть одна образмериваемая точка лежит внутри слоя, отметки высоты – если точка указания лежит внутри слоя, оси строительной подосновы – если их плоскость попала в слой.

Описанное параметрическое представление аксонометрической схемы трубопроводной системы может сохраняться на диске как комплект параметров, без геометрической части. Однако при выборе комплекта параметров на диске для работы с ним пользователь просматривает уже геометрическое изображение АСТС, которое генерируется при перемещении по меню в режиме on-line, как это показано на рис. 1. Следует отметить компактность параметрического представления: комплект параметров показанной на рис.1 схемы занимает на диске всего 2060 байт.

Параметрическое представление АСТС отличается также насыщенностью связей принадлежности. При удалении одной точки удаляются:
- все трубы, подходящие к ней;
- соединения труб, куда входят удаляемые трубы;

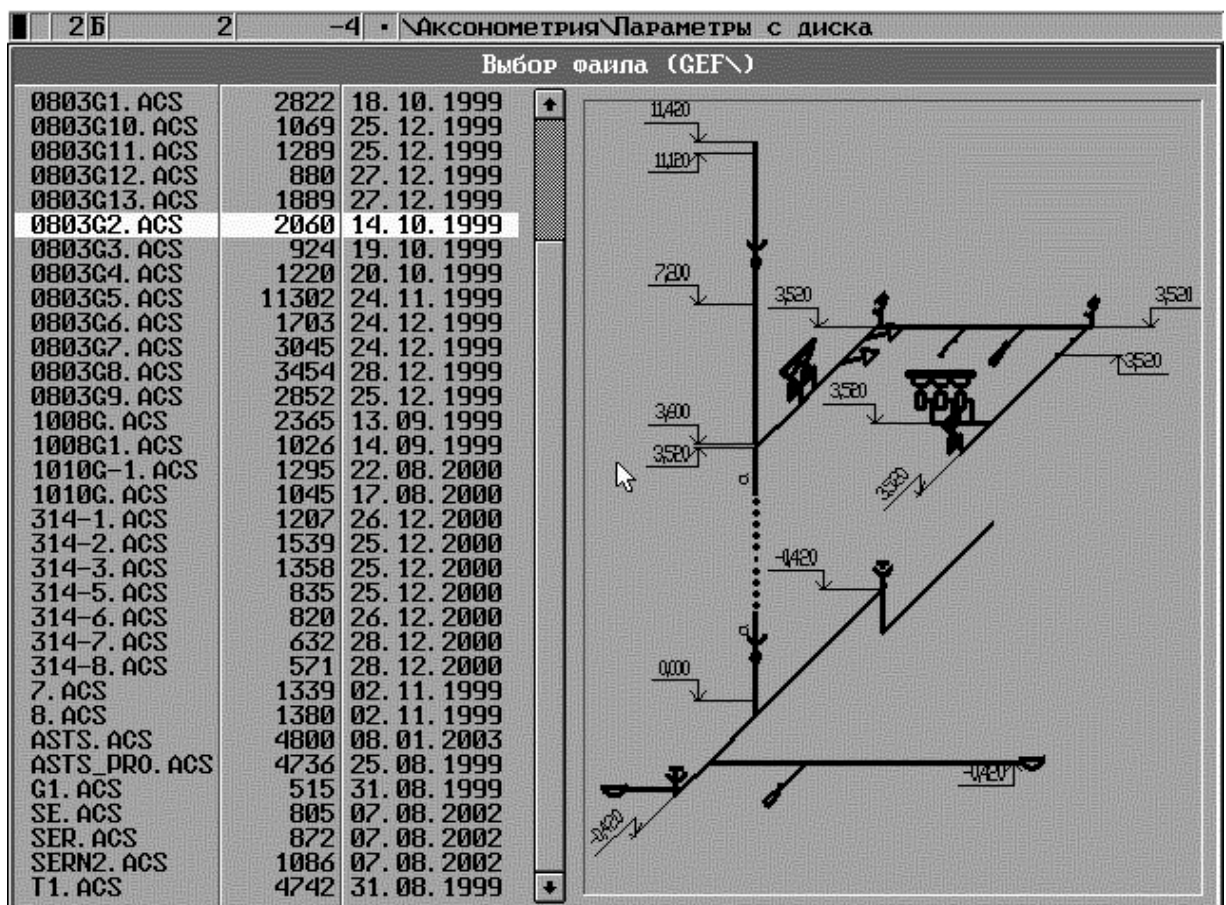

Рис. 1. Выбор комплекта параметров АСТС из имеющихся на диске

- линии разрыва на этих трубах;
- блоки, привязанные только к этим трубам;
- все сноски от текстов на удаляемые трубы и блоки (если удаляется последняя сноска от текста - он также удаляется);
- позиционные обозначения удаляемых труб и блоков (если удаляется последнее позиционное обозначение с таким номером - удаляется и соответствующий комплект специфицирующих свойств);
- все отметки высоты, точки начала выносных линий которых выходят из удаляемых труб и блоков;
- выносные линии в размерах, выходящие из удаляемых точки и блоков (если в размере остается одна выносная линия - он удаляется).

*Использованные источники*